\documentclass{PoS}
\usepackage{graphicx}
\usepackage{fancybox}
\usepackage{fancyhdr}
\usepackage{amsmath}
\usepackage{amssymb}
\usepackage{epsfig}
\usepackage{verbatim}
\title{MC Realization of IR-Improved DGLAP-CS Parton Showers: HERWIRI1.0}

\ShortTitle{MC Realization of IR-Improved DGLAP-CS Parton Showers: HERWIRI1.0}

\author{\speaker{B.F.L. Ward}\thanks{Work partly supported by US DOE grant DE-FG02-09ER41600 and by NATO grant PST.CLG.980342.}\\
        Department of Physics, Baylor University\\
        E-mail: \email{BFL\_Ward@baylor.edu}}

\author{S. Joseph\\
        Department of Physics, Baylor University\\
        E-mail: \email{Sammy\_Joseph@baylor.edu}}
\author{S. Majhi\\
        Theory Division, Saha Institute of Nuclear Physics\\
        E-mail: \email{Swapan.Majhi@saha.ac.in}}
\author{S. A. Yost\\
        Department of Physics, The Citadel\\
        E-mail: \email{Scott.Yost@citadel.edu}}

\abstract{We introduce the new IR-improved 
Dokshitzer-Gribov-Lipatov-Altarelli-Parisi-Callan-Symanzik (DGLAP-CS) kernels recently developed by one of us into
the HERWIG6.5 to generate a new MC, HERWIRI1.0(31),
for hadron-hadron scattering at high energies.
We use MC data to compare the parton shower generated by the 
standard DGLAP-CS kernels and that generated by the new IR-improved 
DGLAP-CS kernels. The seamless
interface to MC@NLO, MC@NLO/HERWIRI, is illustrated.
We show comparisons with FNAL data
and we discuss some possible 
LHC phenomenology implications.\\
}
\FullConference{{\rm BU-HEPP-10-01, Jan., 2010}\\
\vskip.2cm
RADCOR 2009 - 9th International Symposium on Radiative Corrections (Applications of Quantum Field Theory to Phenomenology) ,\\
		 October 25 - 30 2009\\
		 Ascona, Switzerland}

\begin{document}
\newcommand{\mathswitchr}[1]{\relax\ifmmode{\mathrm{#1}}\else$\mathrm{#1}$\fi}
\newcommand{\Ei}{\mathop{\mathrm{Ei}}\nolimits}
\newcommand{\Was}{W\c as}
\newcommand{\FYFS}{F_{\mathrm YFS}}
\newcommand{\rQCED}{\mathswitchr{QCED}}
\baselineskip=13pt
\section{\label{intro} Introduction}

For the era of precision LHC physics($1\%$ or better total theoretcial precision~\cite{jadach1}), we need~\cite{qedeffects,radcor-ew,ditt-lp09}
resummed 
${\cal O}(\alpha_s^2L^n),{\cal O}(\alpha_s\alpha L^{n'}),
{\cal O}(\alpha^2 L^{n''})$ corrections for $n=0,1,2,$ $n'=0,1,2,$ $n''=1,2$, 
in the presence of parton showers, on an event-by-event basis, 
without double counting and with exact phase space.
We present the first step in realizing our 
new MC event generator approach to such precision LHC physics
with amplitude-based QED$\otimes$QCD resummation~\cite{qced} 
by introducing the attendant
new parton shower MC for QCD that follows from our approach. This will set 
the stage for the complete implementation of 
the QED$\otimes$QCD resummed theory in which all IR singularities are
canceled 
to all orders in $\alpha_s$ and $\alpha$.
This new parton shower MC, which is developed 
in the HERWIG6.5~\cite{herwig} environment and which we have
called HERWIRI1.0(31)~\cite{irdglap3-plb}, already shows improvement in
comparison with the FNAL soft $p_T$ data on single $Z$ production
as we quantify below. We also note that, while the 
explicit IR cut-offs in the HERWIG6.5
environment will not be removed here, HERWIRI 
only involves integrable
distributions so that in principle these cut-offs could be removed. 
\par\indent
 
We first review
our approach to resummation and its relationship to those in Refs.~\cite{cattrent,scet}. Section 3 contains a summary of the attendant new IR-improved DGLAP-CS~\cite{dglap,cs} theory~\cite{irdglap1}.
Section 4 presents the implementation of the new IR-improved kernels in the framework of HERWIG6.5~\cite{herwig} to arrive at the new, IR-improved parton shower MC
HERWIRI1.0(31). We illustrate the effects of the IR-improvement 
with the specific
single $Z$ production process at LHC energies. We compare with recent 
data from FNAL to make
direct contact with observation\footnote{From Ref.~\cite{scott1} the current state-of-the-art theoretical precision tag on single $Z$
production at the LHC is 
$(4.1\pm0.3)\%=(1.51\pm 0.75)\%(QCD)\oplus 3.79(PDF)\oplus 0.38\pm 0.26(EW)\%$ 
and the analogous estimate for single $W$ production is $\sim 5.7$\%.}. 
\par\indent 


\section{Review of QED$\otimes$QCD Resummation}
In Refs.~\cite{qced,irdglap1}
we have derived the following expression for the 
hard cross sections in the SM $SU_{2L}\times U_1\times SU_3^c$ EW-QCD theory{\small
\begin{eqnarray}
d\hat\sigma_{\rm exp} &=& e^{\rm SUM_{IR}(QCED)}
   \sum_{{n,m}=0}^\infty\frac{1}{n!m!}\int
\frac{d^3p_2}{p_2^{\,0}}\frac{d^3q_2}{q_2^{\,0}}
\prod_{j_1=1}^n\frac{d^3k_{j_1}}{k_{j_1}} 
\prod_{j_2=1}^m\frac{d^3{k'}_{j_2}}{{k'}_{j_2}}
\nonumber\\
& &  \kern-2cm \times \int\frac{d^4y}{(2\pi)^4}
e^{iy\cdot(p_1+q_1-p_2-q_2-\sum k_{j_1}-\sum {k'}_{j_2})+ D_\rQCED}
\ \tilde{\bar\beta}_{n,m}(k_1,\ldots,k_n;k'_1,\ldots,k'_m),
\label{subp15b}
\end{eqnarray}}
where the new YFS-style~\cite{yfs} residuals
$\tilde{\bar\beta}_{n,m}(k_1,\ldots,k_n;k'_1,\ldots,k'_m)$ have $n$ hard gluons and $m$ hard photons and we illustrate the generic 2f 
final state with momenta $p_2,\; q_2$ specified for
definiteness. The infrared functions ${\rm SUM_{IR}(QCED)},\; D_\rQCED\; $
are defined in Refs.~\cite{qced,irdglap1}. Eq.\ (\ref{subp15b}) 
is exact to all orders in $\alpha$ and in $\alpha_s$.

Given that the approaches in Refs.~\cite{cattrent,scet} to QCD resummation
have been shown in Refs.~\cite{geor1} to be are equivalent and given 
that we show
in Refs.~\cite{irdglap1} that our approach is equivalent to that
in Ref.~\cite{cattrent}, it follows that our approach is also equivalent to that in Ref.~\cite{scet}. See Refs.~\cite{irdglap1} for the attendant
further details.

\section{Review of IR-Improved DGLAP-CS Theory}
The result Eq.\ (\ref{subp15b})
allows us to improve~\cite{irdglap1} in the IR regime 
the kernels in DGLAP-CS~\cite{dglap,cs}
theory as follows, using a standard notation:{\small
\begin{align}
P^{exp}_{qq}(z)&= C_F \FYFS(\gamma_q)e^{\frac{1}{2}\delta_q}\left[\frac{1+z^2}{1-z}(1-z)^{\gamma_q} -f_q(\gamma_q)\delta(1-z)\right],\nonumber\\
P^{exp}_{Gq}(z)&= C_F \FYFS(\gamma_q)e^{\frac{1}{2}\delta_q}\frac{1+(1-z)^2}{z} z^{\gamma_q},\nonumber\\
P^{exp}_{GG}(z)&= 2C_G \FYFS(\gamma_G)e^{\frac{1}{2}\delta_G}\{ \frac{1-z}{z}z^{\gamma_G}+\frac{z}{1-z}(1-z)^{\gamma_G}\nonumber\\
&\qquad +\frac{1}{2}(z^{1+\gamma_G}(1-z)+z(1-z)^{1+\gamma_G}) - f_G(\gamma_G) \delta(1-z)\},\nonumber\\
P^{exp}_{qG}(z)&= \FYFS(\gamma_G)e^{\frac{1}{2}\delta_G}\frac{1}{2}\{ z^2(1-z)^{\gamma_G}+(1-z)^2z^{\gamma_G}\},
\label{dglap19}
\end{align}}
where the superscript ``exp'' indicates that the kernel has been resummed as
predicted by Eq.\ (\ref{subp15b}) when it is restricted to QCD alone and 
where we refer the reader to Refs.~\cite{irdglap1} for the
detailed definitions of the respective resummation functions $\FYFS,\gamma_A,\delta_A,f_G, A=q,G$
\footnote{The improvement in Eq.\ (\ref{dglap19}) 
should be distinguished from 
the also-important
resummation in parton density evolution for the ``$z\rightarrow 0$'' regime,
where Regge asymptotics obtain -- see for example Ref.~\cite{ermlv,guido}. This
latter improvement must also be taken into account 
for precision LHC predictions.}.\par\indent
 
We refer the reader to Refs.~\cite{irdglap1,irdglap4-ichp}
for discussion of 
a number of illustrative results and implications of the new 
kernels and Eq.\ (\ref{subp15b})
that are beyond the scope we have here. The net effect
of the results in Refs.~\cite{irdglap1,irdglap4-ichp}
is that we have a consistent theoretical paradigm based on
Eqs.\ (\ref{subp15b},\ref{dglap19}) for precision LHC theory
that can be systematically improved order-by-order in
perturbation theory with no double counting.
With an eye toward the full MC implementation of our approach, we turn next to the initial stage of that implementation -- that of the new kernels.
\section{Realization of IR-Improved DGLAP-CS Theory via MC Methods}
We have implemented the 
new IR-improved kernels in the HERWIG6.5 environment to produce
a new MC, HERWIRI1.0, which stands for ``high energy radiation with IR improvement''\cite{bw-ms-priv-a}.\par\indent

Specifically, we modify the kernels in the HERWIG6.5 module HWBRAN and in the attendant
 related modules\footnote{We thank M. Seymour and B. Webber for helpful discission.} with the following substitutions: ~{\small
$\text{DGLAP-CS}\; P_{AB}  \Rightarrow \text{IR-I DGLAP-CS}\; P^{exp}_{AB}$}
while leaving the hard processes alone for the moment. We have in 
progress~\cite{inprog}
the inclusion in our framework of YFS synthesized electroweak  
modules from Refs.~\cite{jad-ward}
for
HERWIG6.5, HERWIG++~\cite{herpp}, and MC@NLO~\cite{mcnlo} hard processes\footnote{Similar results for PYTHIA~\cite{pythia} and for the new kernel evolution in Ref.~\cite{skrzjad} are under study.}, 
as the
CTEQ~\cite{cteq} and MRST(MSTW)~\cite{mrst} best (after 2007) parton densities
do not include the precision electroweak higher order corrections that are needed in a 1\% precison tag budget for processes such as single heavy gauge boson production in the LHC environment~\cite{radcor-ew}. 
\par
\begin{figure}[h]
\begin{center}
\includegraphics[height=80mm,angle=90]{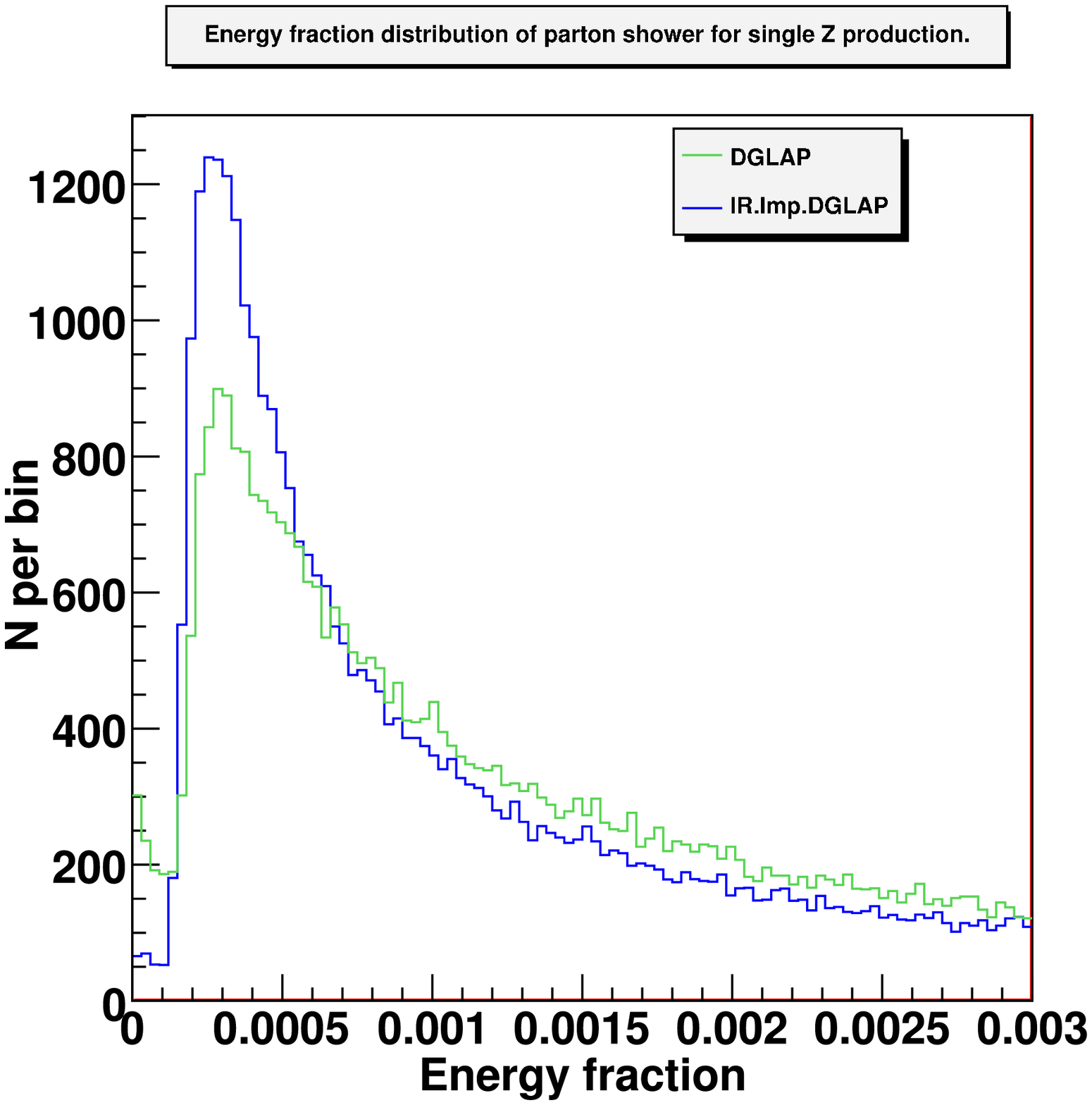}
\end{center}
\caption{The $z$-distribution(ISR parton energy fraction) shower comparison in HERWIG6.5.}
\label{fighw5}
\end{figure}

\begin{figure*}[t]
\centering
\setlength{\unitlength}{0.1mm}
\begin{picture}(1600, 830)
\put( 420, 800){\makebox(0,0)[cb]{\bf (a)} }
\put(1200, 800){\makebox(0,0)[cb]{\bf (b)} }
\put(   -90, 0){\makebox(0,0)[lb]{\includegraphics[height=86mm,angle=90]{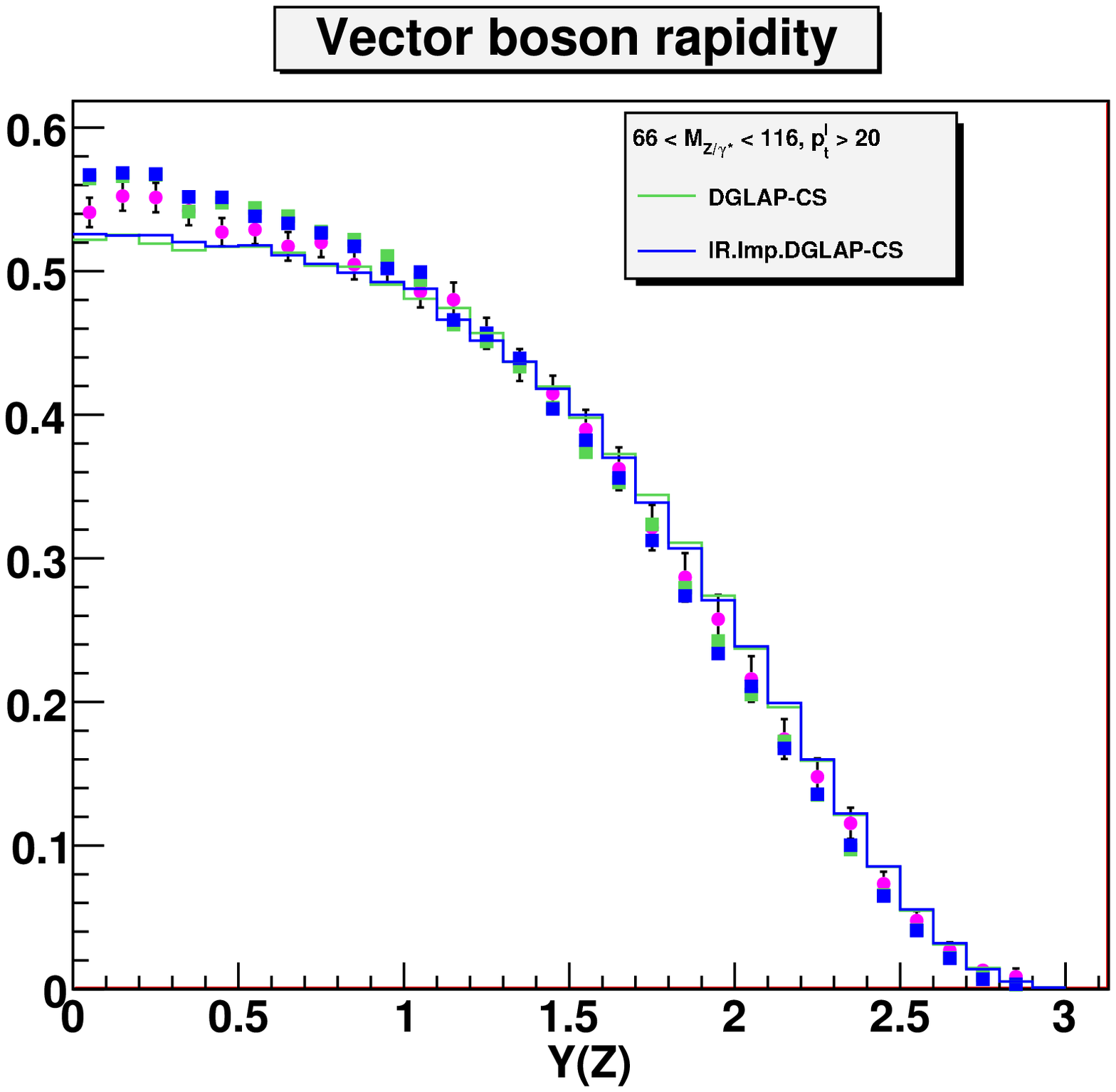}}}
\put( 770, 0){\makebox(0,0)[lb]{\includegraphics[height=86mm,angle=90]{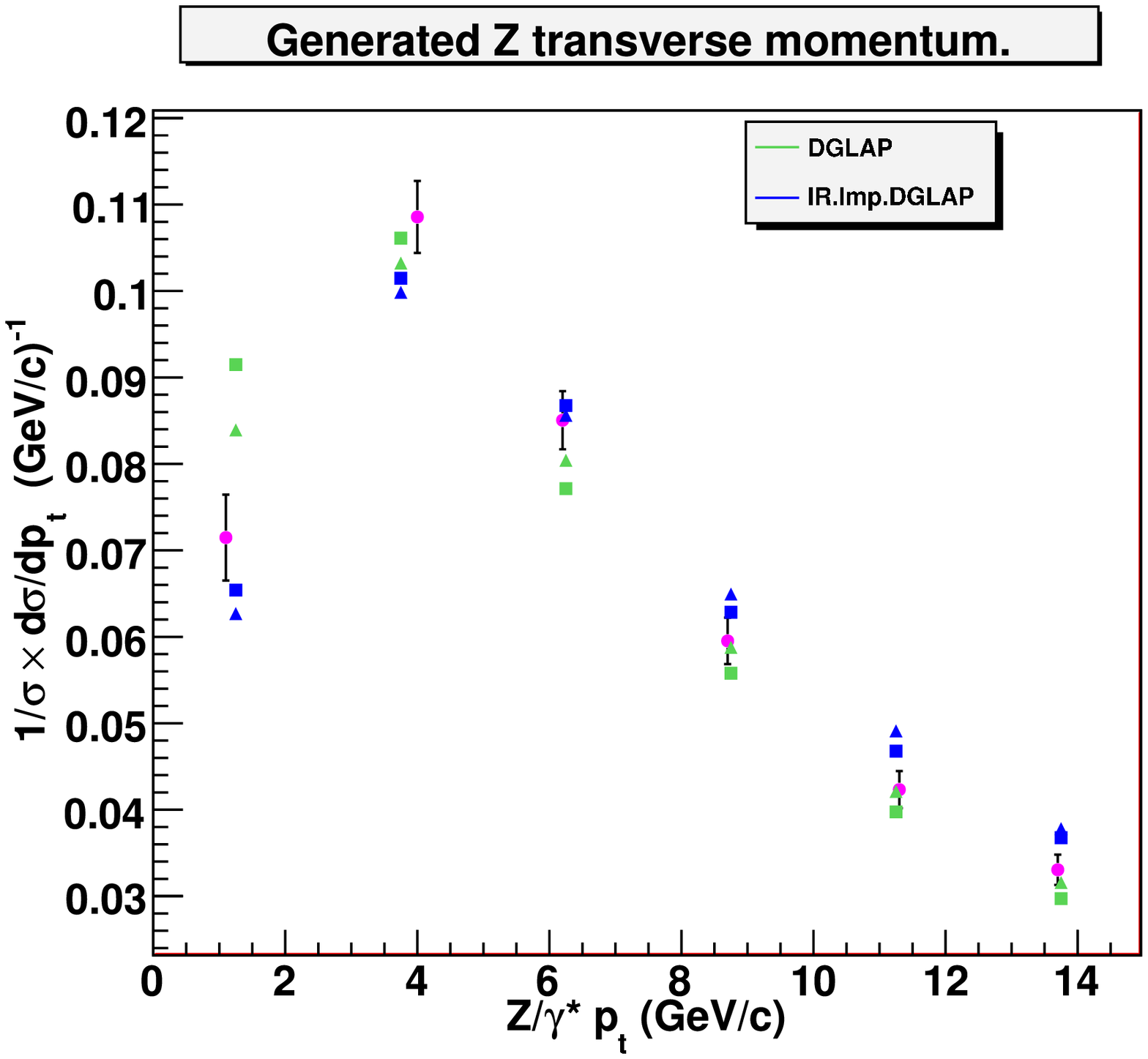}}}
\end{picture}
\caption{Comparison with FNAL data: (a), CDF rapidity data on
($Z/\gamma^*$) production to $e^+e^-$ pairs, the circular dots are the data, the green(blue) lines are HERWIG6.510(HERWIRI1.031); 
(b), D0 $p_T$ spectrum data on ($Z/\gamma^*$) production to $e^+e^-$ pairs,
the circular dots are the data, the blue triangles are HERWIRI1.031, the green triangles are HERWIG6.510 -- in both (a) and (b) the blue squares are MC@NLO/HERWIRI1.031, and the green squares are MC@NLO/HERWIG6.510. These are untuned theoretical results.
}
\label{fighw9}
\end{figure*}

\indent
The details of the implementation are given in Refs.~\cite{irdglap3-plb,irdglap4-ichp} and we do not reproduce them here due to a lack
of space. We have done many comparisons of the properties of the
parton showers from HERWIG6.510 and HERWIRI1.031. In general, as we
show here in Fig.~\ref{fighw5},
the IR-improved showers tend to be softer
in the energy fraction variable $z=E/E_{Beam}$ where $E\; (E_{Beam})$ is the cms parton(beam)
energy for hadron-hadron scattering respectively.
See Refs.~\cite{irdglap3-plb,irdglap4-ichp} for the complete
discussion of such comparisons. We have also made comparison 
analyses with the data from 
FNAL on the $Z$ rapidity and $p_T$ spectra as reported in 
Refs.~\cite{galea,d0pt}. We show these results, for $1.96$TeV cms 
energy, in Fig.~\ref{fighw9}. We see that HERWIRI1.0(31)
and HERWIG6.5 both give a reasonable 
overall representation of the CDF rapidity data but that
HERWIRI1.031 is somewhat closer to the data for small values of $Y$. The two $\chi^2$/d.o.f are 1.77 and 1.54
for HERWIG6.5 and HERWIRI1.0(31) respectively. The data 
errors in Fig.~\ref{fighw9}(a)
do not include luminosity and PDF errors~\cite{galea}, so that
they can only be used conditionally at this point. We note as well that 
including the NLO
contributions to the hard process via MC@NLO/HERWIG6.510
and MC@NLO/HERWIRI1.031\cite{mcnlo}\footnote{We thank S. Frixione for
helpful discussions with this implementation.} improves the agreement for both
HERWIG6.510 and for HERWIRI1.031, where the $\chi^2$/d.o.f are changed
to 1.40 and 1.42 respectively.
That they are both consistent with
one another and within 10\% of the data in the low $Y$ region is
fully consistent with what we expect given
our comments about the errors and the generic accuracy of
an NLO correction in QCD. A more precise discussion at the NNLO
level with DGLAP-CS IR-improvement and
a more complete discussion  of the errors will
appear~\cite{elswh}. These rapidity
comparisons are then
important cross-checks on our work. 
We also see that HERWIRI1.031 gives a better fit to
the D0 $p_T$ data
compared to HERWIG6.510 for low $p_T$, 
(for $p_T<12.5$GeV, the $\chi^2$/d.o.f. are
$\sim$ 2.5 and 3.3 respectively if we add the statistical and systematic
errors), showing that the IR-improvement makes a better representation
of QCD in the soft 
regime for a given fixed order in perturbation theory.
We have also added
the results of MC@NLO~\cite{mcnlo}
for the two programs and we see
that the ${\cal O}(\alpha_s)$ correction improves the $\chi^2$/d.o.f for
the HERWIRI1.031 in both the soft and hard regimes and it improves
the HERWIG6.510 $\chi^2$/d.o.f for $p_T$ near $3.75$ GeV
where the distribution peaks. For $p_T<7.5$GeV the $\chi^2$/d.o.f for
the MC@NLO/HERWIRI1.031 is 1.5 whereas that for MC@NLO/HERWIG6.510 is somewhat
worse.
These results are of course still subject to tuning as we indicated above. 
We await further tests of the new
approach, both at FNAL and at LHC.
\par
\begin{acknowledgments}
%
One of us (B.F.L.W) acknowledges helpful discussions with Prof. Bryan Webber
and Prof. M. Seymour and with Prof. S. Frixione. B.F.L. Ward also thanks Prof. L. Alvarez-Gaume and Prof. W. Hollik for the support and kind hospitality of the CERN TH Division and of the Werner-Heisenberg Institut, MPI, Munich, respectively, while this work was in progress. S. Yost acknowledges the hospitality and support of Princeton University and a grant from The Citadel Foundation.

\end{acknowledgments}

\end{document}